\documentclass[a4paper]{mem}
\usepackage{natbib}
\usepackage{graphicx}
\usepackage[a4paper]{hyperref}
\idline{73}{23}
\begin{document}
\title{The XMM--LSS project: a short presentation of the survey and of the first results}

   \author{S. Andreon \inst{1}, 
   M. Pierre \inst{2}\\ 
           for the XMM--LSS collaboration 
}
   \offprints{S. Andreon}

   \institute{INAF--Osservatorio Astronomico di Brera, Italy, \\
\email{andreon@brera.mi.astro.it}\\ 
              \and  
	CEA/DSM/DAPNIA, Service d'Astrophysique, Saclay, France      
             }

   \abstract{

We (Pierre et al. 2003) have designed a medium deep large area X-ray 
survey with XMM -- the XMM Large Scale Structure survey, XMM-LSS  --
with the scope of extending the cosmological tests attempted using ROSAT
cluster samples to two redshift bins between $0<z<1$ while maintaining
the precision of earlier studies.  The optimal survey design was found
to be an $8\degr\times 8\degr$ area, paved with 10 ks XMM pointing
separated by 20 arcmin. This area is the target of several 
complementary surveys, from
the ultraviolet to the radio wavelengths, allowing, beside cosmological
studies, detailed studies of the objects included in our area. 

   }
   \authorrunning{Andreon et al.}
   \titlerunning{The XMM--LSS project: first results}
   \maketitle
%
%________________________________________________________________

\newcommand{\gtapprox}{\raisebox{-0.5ex}{$\,\stackrel{>}{\scriptstyle\sim}\,$}}

\begin{figure*}
\centering
\vspace{4cm}
\resizebox{\hsize}{!}{\includegraphics{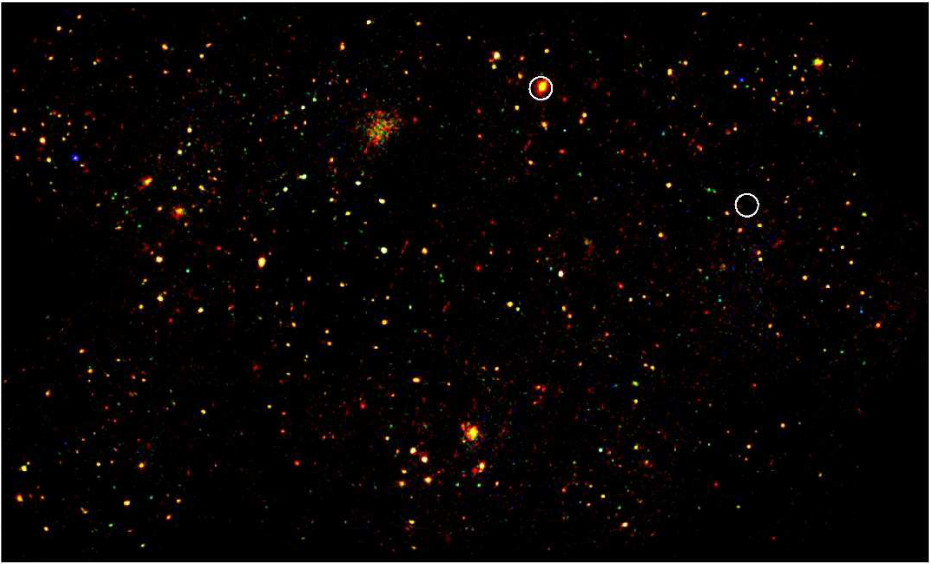}}
\caption{{\bf First view of the deep X-ray sky on large
scales}. Image obtained combining the first 15 XMM-LSS fields
mosaiced in true X-ray colours: red [0.3-1.0] keV, green
[1.0-2.5] keV, blue [2.5-10.0] keV. The circles indicate the
sources found in the RASS; the brightest one being a star,
HD14938. The displayed region covers 1.6 deg$^2$. This is the first
time that such an X-ray depth has been achieved over such an
area. The improvement with respect to the RASS is striking, with
a source density of the order of $\sim$ 300~deg$^{-2}$ in the
[0.5-2] keV band. The wealth of sources includes supersoft and
very hard sources, as well as sources with a wide range of
intrinsic extents, giving an indication of the scientific
potential of the XMM-LSS survey.  The extended source at the
top is a cluster of galaxy.
}
\end{figure*}

\section{Introduction} 

From a simple theoretical point of view, clusters
of galaxies - the most massive bound structures in the Universe - are
objects having a mass of the order of $10^{14-16}~ M_\odot$ growing by
accretion at a rate governed by the initial density fluctuation
spectrum, the cosmological parameters, the nature and amount of dark
matter as well as the nature of the dark energy. Their 3-dimensional
space distribution and number density as functions of cosmic time
constrain cosmological parameters in a unique way. Clusters offer considerable
advantages for large scale structure (LSS) studies: they can provide
complete samples of objects over a very large volume of space, and they
are in crucial respects simple to understand. The extent (and mass) of
clusters can be traced by their X-ray emission while the theory
describing their formation (biasing) and evolution from the initial
fluctuations can be tested with numerical simulations. Such a level of
understanding does not exist for galaxies - which have reached a highly
non-linear stage - to and even less for QSO and AGN formation. The
resulting cluster LSS counts studies can constrain cosmological
parameters, independently of Cosmic Microwave Background (CMB) and
supernova (SN) studies since they do not rely on the same physical
processes. A quantitative overview of the cosmological implications of
cluster surveys can be found for instance in Haiman, Mohr \& Holder
(2001). 

XMM is in a position to open a new era for X-ray surveys. Its high
sensitivity, considerably better PSF than the RASS ($6\arcsec$ on axis)
and large field of view ($30\arcmin$), make it a powerful tool for the
study of extragalactic LSS. In this respect, two key points may be
emphasized. Firstly, a high galactic latitude field observed with XMM at
medium sensitivity ($\sim 0.5-1~ 10^{-14}$ erg~cm$^{2}$~s$^{-1}$) is
``clean'' as it contains only two types of objects, namely QSOs
(pointlike sources) and clusters (extended sources) well above the
confusion limit. Secondly, if clusters more luminous than $L_{[2-10]}
\sim 3 ~10^{44}$ h$_{70}^{-2}$erg/s are present at high redshift, they
can be detected as extended sources out to $z = 2$, in XMM exposures of
only 10 ks.

\section{The XMM--LSS survey and the associate surveys}
 
We (Pierre et al. 2003) have designed a survey to yield some 800
clusters in  two redshift bins with $0<z<1$: the XMM Large Scale
Structure Survey (XMM-LSS). The optimal survey design was found to be an
$8\degr\times 8\degr$ area, paved with 10 ks XMM pointing separated by
20 arcmin (i.e. 9 pointings per degree square). The expected ultimate
sensitivity is $\sim 3~10^{-15}$~erg~cm$^{2}$~s$^{-1}$ for pointlike
sources in the [0.5-2] keV band. It will also trace the LSS as defined
by X-ray QSOs out to redshifts of $\sim 4$.

In addition, the proposed X-ray survey is associated with several other
major new generation surveys (optical, IR, Radio, UV), as detailed in
Table 1.

\begin{figure}
\centering
\vspace{4cm}
\includegraphics[width=6.5truecm,clip]{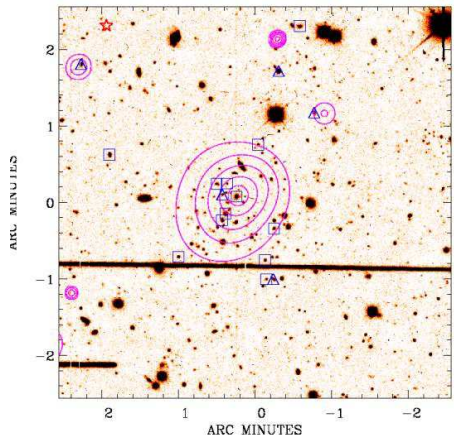}
\caption{X--ray contours
superposed to an I band image of a clusters at $z=0.84$. Marked
galaxies are spectroscopic confirmed members. The cluster velocity
dispersion, using 17 member galaxies, is $\sim800$ km/s. The dark
horizontal lines are an artifact due to a nearby bright star.}
\end{figure}

\begin{figure}
\centering
\vspace{4cm}
\includegraphics[width=6.5truecm]{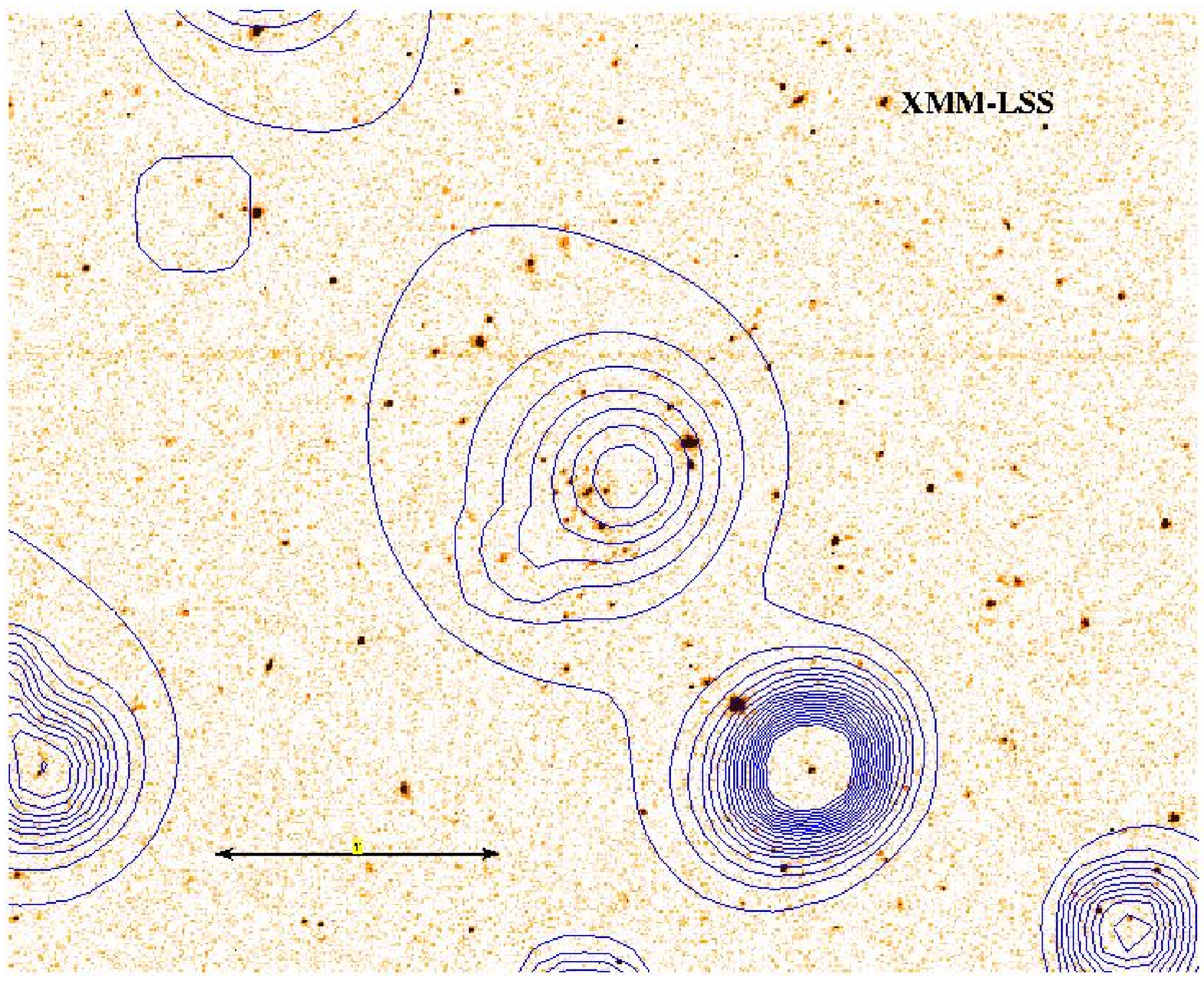}
\caption{X-ray contours (wavelets filtered) superposed to the
$K$-band image of one of XMM--LSS clusters at $z\sim1.0$. The
photometric redshift of the cluster is derived from the $R-K$ color of
the cluster color--magnitude relation.} 
\end{figure}

\section{First results}

During fall 2002, we did perform near--infrared observations at NTT of
$z\gtapprox 1$ clusters and trial spectroscopic observations of
candidate clusters at VLT and Magellan telescopes. Figure 2 shows  an
example of a cluster spectroscopically confirmed at $z=0.8$
(Valtchanov et al. 2003).  Spectroscopic  observations confirm 17 out 18
of the candidate clusters observed (Valtchanov et al. 2003, Willis et al.
2003). 

Candidate $z\gtapprox1$ clusters are extracted from XMM--LSS database as
extended X-ray sources with no obvious optical counterpart (our optical
data are deep enough to detect clusters up to $z\sim1$) and observed in
the near--infrared using SOFI at NTT (Andreon et al. 2003a).  The run was
very useful, confirming 5 $z\gtapprox 1$ clusters in just over 1 deg$^2$
of the survey and suggesting a few $z\gtapprox 1.3$ clusters. One of
the near--infrared confirmed clusters is shown in Figure 3.

\begin{table*}
\centering
\begin{tabular}{|l|l|l|l|}
\hline
Observatory/Instrument & (Planned) Coverage & Band & Final
Sensitivity \\ 
\hline
XMM/EPIC & 64 deg$^2$ & [0.2-10] keV & $\sim 3~10^{-15}$erg~cm$^{2}$~s$^{-1}$ [1] \\
CFHT/CFH12K (VVDS Deep) *& 2 deg$^2$\ GT & B, V, R, I & 26.5, 26.0, 26.0, 25.4 [2] \\
CFHT/CFH12K (VVDS Wide) *& 3 deg$^2$\ GO & V, R, I & 25.4, 25.4, 24.8 [2] \\
CFHT/MegaCam & 72 deg$^2$ & u*, g', r', i', z' & 25.5, 26.8,26.0, 25.3, 24.3 [3] \\
CTIO 4m/Mosaic & $\sim$ 16 deg$^2$ & R, z' & 25, 23.5 [4] \\
UKIRT/WFCAM & 8.75 deg$^2$ & J, H, K& 22.5, 22.0, 21.0 [5] \\
VLA/A-array *& 110 deg$^2$ & 74 MHz & 275 mJy/beam [6a]\\
VLA/A-array & 5.6 deg$^2$ & 325 MHz & 4 mJy/beam [6b]\\
OCRA & all XMM-LSS clusters & 30 GHz & 100 $\mu$Jy [7]\\
AMiBA & 70 deg$^2$ & 95 GHz & 3.0 mJy [8] \\
SIRTF/IRAC (SWIRE Legacy) & 8.7 deg$^2$ & 3.6, 4.5, 5.8, 8.0 $\mu$m &
7.3, 9.7,27.5, 32.5 $\mu$Jy [9a]\\
SIRTF/MIPS (SWIRE Legacy) & 8.9 deg$^2$ & 24, 70, 160 $\mu$m & 0.45,
6.3, 60 mJy [9b]\\
Galex & $\sim 20$ deg$^2$ & 1305-3000 \AA & $\sim 25.5$ [10] \\
\hline\hline
\end{tabular} 
\caption[]{{\bf XMM-LSS X-ray and associated surveys} Notes: * :
complete[1]: for pointlike sources in [0.5-2] keV[2]: AB$_{Mag}$,
$5\arcsec$ aperture[3]: S/N = 5 in $1.15\arcsec$ aperture[4]: 4
sigma in $3\arcsec$ aperture[5]: Vega$_{Mag}$ [6a]: $30\arcsec$
resolution; deeper observations planned[6b]: $6.3\arcsec$
resolution[7] $5 \sigma$, detection limit[8] $6 \sigma$, detection
limit[9a] $5 \sigma$[9b] $5 \sigma$[10]: AB$_{Mag}$
\label{flwp}}
\end{table*}

The wealth of data available for almost a thousand of 
clusters allows to study the properties of cluster galaxies, their
evolution (both with lookback time and during the infall in the
cluster), while keeping the different galaxies classes separated. 
At the time of this writing, the accumulated sample is already the 
largest at high redshift ever studied. In a
first exploratory study, we focus on the evolution of both the reddest
galaxies and of the whole cluster galaxy population. Due to space
limitations, we summarize results for only the latter. Figure 4 shows
the Schechter (1976) characteristic magnitude, $m^*$, in the $z'$ band
($\lambda\sim9000$ \AA), for some of our clusters.  
The upper curve shows the expected relationship between redshift and $m^*$
if stars do not evolve, and it is
clearly rejected by the data. A model in which stars form at $z_f=5$
(middle curve) reproduces well the faintest $m^*$ (at a given redshift,
id est the curve traces well the upper envelope of the data points).
Clusters with $m^*$ brighter than the prediction of a $z_f=1.3$ model
(bottom curve), show a re-youngening, due to a secondary star--formation
activity happened at ``low" redshift, maybe related to the Butcher--Oemler
effect (Butcher \& Oemler 1984). The reader is invited to refer to
Andreon et al. (2003b) for details.

\section{Conclusion}
The first optical and spectroscopic observations of the XMM--LSS
fields observed during the AO-1 period show the feasibility
of the cosmological XMM--LSS program, and give interesting
results on the galaxy evolution up to $z=1$.

\begin{figure}
\centering
\includegraphics[width=6.5truecm]{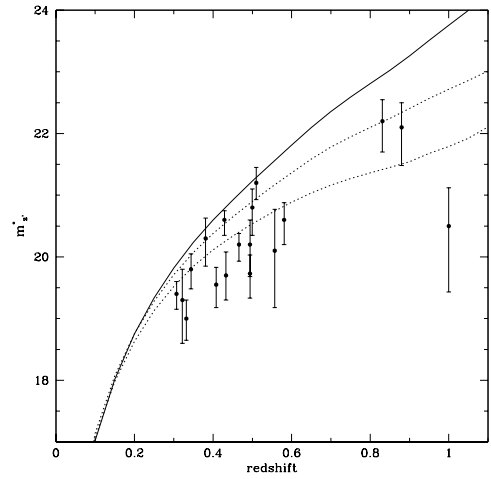}
\caption{Evolution of M* in the $z'$ filter (points) and
model predictions (curves).
Three models are plotted in the figure:
a model without star aging (top curve), 
and two models with formation redshifts $z_f=5$ (middle curve)
and $z_f=1.3$ (bottom curve).} 
\end{figure}

%\begin{acknowledgements}
% We acknowledge J. W. Bush, for given a new sense to the word `peace'.
%\end{acknowledgements}

\bibliographystyle{aa}

\begin{thebibliography}{}

\bibitem[]{} 
Andreon S. et al, 2003a,  {\em in preparation}

\bibitem[]{} 
Andreon S. et al, 2003b, {\em in preparation}

\bibitem[Butcher \& Oemler(1984)]{1984ApJ...285..426B} 
Butcher, H.~\& Oemler, A.\ 1984, ApJ, 285, 426 

\bibitem[{Haiman et al }{2001}]{hai}
Haiman Z., Mohr J. J., Holder G. P., 2001, ApJ, 553, 545

\bibitem[]{}
Pierre et al. 2003, A\&A, submitted (astro-ph/0305191)

\bibitem[Schechter(1976)]{1976ApJ...203..297S} 
Schechter, P.\ 1976, ApJ, 203, 297 

\bibitem[]{}
Valtchanov I. et al 2003, A\&A, submitted (astro-ph/0305192)

\bibitem[]{}
Willis et al. 2003, {\em in preparation}

\end{thebibliography}

\end{document}